\documentclass[twocolumn]{article}
\usepackage{pkuarticle}

\usepackage{algorithmic}
\usepackage{comment}
\usepackage{textcomp}
\usepackage{multirow}
\usepackage{array}
\usepackage{upgreek}
\usepackage{enumerate}

\usepackage[numbers,sort&compress]{natbib}

\definecolor{Bittersweet}{RGB}{254,111,94}
\definecolor{ForestGreen}{RGB}{0,155,58}

\newcommand{\rev}[1]{\textcolor{black}{#1}}

\title{Optical Flow Sensor: A Direction-Selective Bionic Retina Design}

\author{Juchen Zhou, Bonan Yan*, Yuchao Yang*}

\affiliation{%
  Peking University, China\\
  \small Correspondence to \texttt{bonanyan@pku.edu.cn}%
}

\begin{document}

\maketitle

\begingroup
\renewcommand{\thefootnote}{}%
\footnotetext{\footnotesize
  This paper has been accepted for publication for \textit{Computer}. \copyright\ IEEE.
  Personal use of this material is permitted. Permission from IEEE must be
  obtained for all other uses, in any current or future media, including
  reprinting/republishing this material for advertising or promotional
  purposes, creating new collective works, for resale or redistribution to
  servers or lists, or reuse of any copyrighted component of this work in
  other works. Any work that develops, builds upon, or is substantially
  similar to the optical flow sensor design presented herein should cite
  this work or the published version in \textit{Computer}.}%
\endgroup

\begin{abstract}
Optical flow characterizes motion in the visual field and is fundamental to motion perception and tracking in biological and artificial vision systems. Biological retinas extract motion efficiently through local ON/OFF pathways and parallel processing, while conventional frame-based optical flow relies on dense sampling and global computation, resulting in high latency and power consumption. To overcome these limitations, we present a pixel-level Optical Flow Sensor (OFS) integrated circuit. The design combines Dynamic Vision Sensor (DVS) ON/OFF event comparison with time-difference measurement to enable fully parallel optical flow computation on-chip. An optical-flow-specific Address-Event Representation (OF-AER) interface supports low-power, high-throughput readout. \rev{Based on the CMOS-based OFS, we further propose optical memristor-based OFS to reduce sensor power consumption and area overhead.} Experimental results show that the proposed OFS achieves a 303$\times$ reduction in power consumption compared with FPGA-accelerated DVS systems while maintaining microsecond-level latency. Moreover, by directly outputting optical flow vectors, the OFS reduces output data size by approximately 3.3$\times$, demonstrating strong potential for ultra-high-speed, low-power vision sensing applications.
\end{abstract}

\noindent\textbf{Keywords:} optical flow sensor, dynamic vision sensor, bionic retina, address-event representation, neuromorphic computing, optical memristor

\section{Introduction}

Motion perception has long been a central topic in biological vision research, where controlled neurophysiological experiments have revealed the computational strategies used by living organisms to detect and interpret movement. Classic behavioral and electrophysiological studies in insects demonstrated the effectiveness of the Hassenstein--Reichardt correlator~\cite{HassensteinReichardt+1956+513+524}, in which delayed and undelayed photoreceptor signals interact to generate direction-selective responses. Similarly, experiments in mammalian retinas uncovered the Barlow--Levick motion detection mechanism~\cite{https://doi.org/10.1113/jphysiol.1965.sp007638}, showing how specific inhibitory circuits contribute to directional selectivity in rabbit retinal ganglion cells. More recent biological experiments in flies and primates~\cite{wang2023type} have confirmed the existence of direction-selective cells. These biological experiments collectively provide strong empirical evidence for the principles underlying motion estimation and have inspired numerous computational frameworks used in artificial vision systems.

Meanwhile, computer scientists adopted the abstract notion of motion perception in the form of optical flow, which describes how pixel intensities evolve across time within a vision field. Optical flow serves as a cornerstone for numerous tasks, ranging from motion perception~\cite{hlxuechip2401} and visual tracking~\cite{7989517} to depth inference~\cite{Hadviger16022021} and scene understanding~\cite{7833065}. Each optical flow vector characterizes the displacement of a pixel between successive observations, while the motion field describes the larger-scale structure of scene dynamics. For autonomous robots, drones, and vehicles, precise optical flow with minimal latency remains indispensable.

\begin{figure}[!t]
	\centering
	\includegraphics[width=\columnwidth]{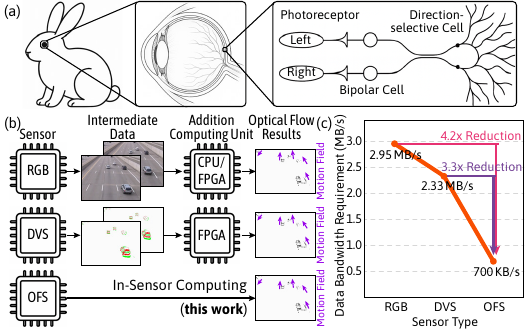}
	\caption{(a) Schematic of the rabbit retina. (b) Several methods to compute optical flow. (c) Out-of-sensor data bandwidth requirements.}
	\label{fig:1}
\end{figure}

Despite its centrality, conventional optical flow estimation is tightly coupled to frame-based imaging pipelines. These pipelines accumulate dense intensity images and then apply algorithmic procedures to infer motion, resulting in excessive data redundancy and substantial latency (Figure~\ref{fig:1}). Dynamic vision sensors (DVS) introduced an alternative paradigm by mimicking retinal spike-based signaling: only brightness changes trigger events, with temporal precision on the order of microseconds~\cite{4444573}. This representation dramatically reduces bandwidth demands and motion blur. However, like biological photoreceptors that require circuitry to produce motion selectivity, DVS do not generate motion cues. Optical flow extraction still relies on external processors or FPGA accelerators, which increases power consumption and constrains system responsiveness.

Optical flow estimation techniques can be classified into three primary categories. Conventional frame-based approaches, such as Horn--Schunck~\cite{HORN1981185} and Lucas--Kanade~\cite{10.5555/1623264.1623280}, provide mathematical frameworks for motion analysis but necessitate solving complex optimization tasks that require substantial computational resources. Contemporary deep learning architectures, including PWC-Net~\cite{sun2018pwcnetcnnsopticalflow} and RAFT~\cite{10.1007/978-3-030-58536-5_24}, deliver enhanced motion estimation precision yet inflicting higher processing power and memory capacity. In contrast, event-based optical flow methodologies, implementing techniques like block-matching~\cite{8050295} and plane-fitting~\cite{8351588}, bring computational processing nearer to the sensor array while still executing motion analysis externally. This architectural separation between data acquisition and processing introduces communication overheads that conflict with the energy-efficient information processing mechanisms observed in biological visual systems.

Inspired by the distributed near-sensor processing in natural vision systems, we present a novel bionic retina design, Optical Flow Sensor (OFS), that carries out motion extraction directly within its pixel array (\rev{Figure~\ref{fig:1}(b)}). \rev{In the CMOS-based OFS, photodiodes are employed as the light-sensing elements to trigger event-driven computation.} Each pixel incorporates minimal circuitry capable of comparing event timestamps and computing local motion cues, enabling the entire array to detect optical flow with massive parallelism and extremely low latency. To facilitate efficient data transmission, we also develop a streamlined, \rev{optical-flow-specific Address-Event Representation} (OF-AER) interface that generates sparse, event-based motion signals, eliminating the need for full-frame reconstruction or the transfer of complete event streams. \rev{Further, we propose the OFS utilizing optical memristors. This approach significantly reduces pixel area and power consumption compared to traditional CMOS-based OFS.}

Our experimental results indicate that the OFS delivers flow estimation accuracy on par with existing methods while consuming 303$\times$ less power, achieving microsecond-scale latency, and reducing the output data rate by 3.3$\times$ relative to conventional DVS--processor setups. These findings confirm that performing optical flow computation directly on the sensor is not only viable but also offers substantial benefits, thereby enabling ultra-efficient vision architectures and advancing next-generation edge-intelligent systems.

\section{Preliminaries}

\subsection{Dynamic Vision Sensors}

DVS represents a neuromorphic sensing paradigm that fundamentally differs from conventional frame-based cameras. Instead of capturing periodic full-frame images, DVS generates discrete events exclusively when significant temporal intensity changes occur at individual pixels. This bio-inspired approach delivers microsecond-level temporal resolution and an exceptional dynamic range exceeding 120 dB, making it particularly suitable for applications requiring high-speed processing, extreme illumination variations, or energy efficiency, such as robotic vision systems and edge computing devices.

At the pixel level, DVS continuously monitors logarithmic intensity variations. Each pixel operates independently and asynchronously, triggering an event when the accumulated brightness change surpasses a contrast threshold $C$:
\begin{equation}
	|\Delta L| = |\ln I(t) - \ln I(t_{last})| \geq C.
\end{equation}
Here, $I$ represents incident brightness, and $t_{last}$ is the timestamp of the last event from the same pixel. Events are polarity-coded: brightness increases generate ON events, while decreases produce OFF events. This event-based representation provides a highly efficient spatiotemporal description of scene dynamics, minimizing data redundancy while preserving critical motion information.

\subsection{Optical Flow}

Optical flow describes the motion field resulting from relative movement between a visual sensor and the scene, as captured on the image plane. It is defined as the instantaneous velocity of projected points, given by \(\boldsymbol{v} = \frac{d\boldsymbol{r}}{dt}\), where \(\boldsymbol{r}\) represents image coordinates. A common foundation for optical flow estimation is the brightness constancy assumption, which posits that image intensity remains constant over small time intervals:
\begin{equation}
	I(\boldsymbol{r},t) = I(\boldsymbol{r} + d\boldsymbol{r}, t + dt).
	\label{eq:BCA}
\end{equation}
Applying a first-order Taylor expansion yields a differential constraint linking intensity gradients to the flow vector:
\begin{equation}
	\begin{aligned}
		&\nabla I(\boldsymbol{r},t) \cdot \boldsymbol{v}(\boldsymbol{r},t) + \frac{\partial I(\boldsymbol{r},t)}{\partial t} = \\ 
		&\frac{\partial I}{\partial x}v_x + \frac{\partial I}{\partial y}v_y + \frac{\partial I}{\partial t} = 0.
		\label{eq:OFE}
	\end{aligned}
\end{equation}
This constraint alone is under-constrained, leading to the aperture problem where the 2D motion vector cannot be uniquely determined. To resolve this, classical methods incorporate additional priors, such as global smoothness in the Horn--Schunck approach~\cite{HORN1981185} or local consistency in Lucas--Kanade~\cite{10.5555/1623264.1623280}. More recent techniques, including block-matching~\cite{8050295} and deep learning models like PWC-Net~\cite{sun2018pwcnetcnnsopticalflow} and RAFT~\cite{10.1007/978-3-030-58536-5_24}, achieve high accuracy but often require significant computational resources, limiting real-time on-chip implementation due to latency and energy concerns. To overcome these issues, we propose a pixel-parallel optical flow sensor (OFS) that enables efficient, low-latency motion estimation directly at the sensor level.

\section{Principle For Designing OFS}
\label{principle}

\begin{figure*}[!t]
	\centering
	\includegraphics[width=\textwidth]{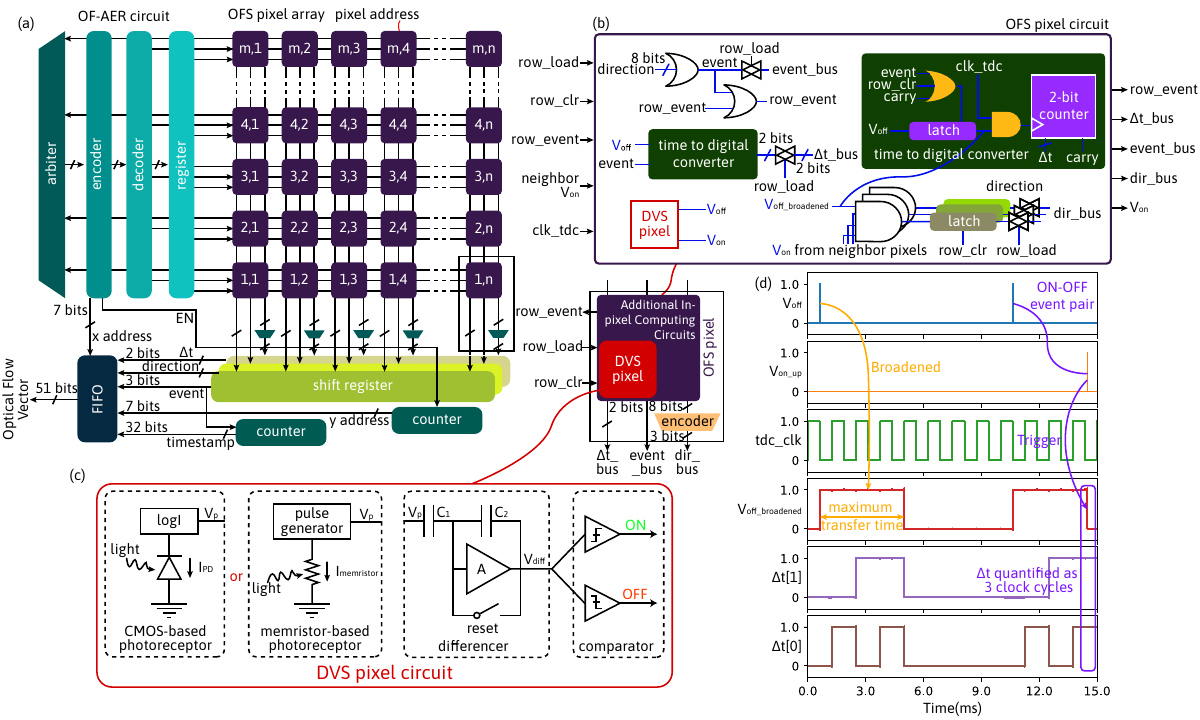}
	\caption{(a) Circuit scheme of the proposed OFS. (b) OFS pixel circuit schematic. (c) DVS pixel implemented with photodiode/optical memristor. \rev{(d) Timing diagram of the TDC circuit.}}
	\label{fig:arch}
\end{figure*}

This section transforms the classical optical flow constraint into a discrete formulation suitable for pixel-based sampling and local measurement evaluation. The primary goal is to derive a compact velocity representation dependent solely on intensity variations between a pixel and its neighbors within a brief temporal interval. Starting from the optical flow equation in Eq.~\ref{eq:OFE}, we approximate spatial derivatives using finite differences along direction $\Delta\hat{\boldsymbol{r}}$ and temporal derivatives over interval $\Delta t$, yielding the discrete optical flow equation:

\begin{equation}
	\frac{I(\boldsymbol{r}+\Delta \boldsymbol{r},t)-I(\boldsymbol{r},t)}{ \Delta \boldsymbol{r} }\Delta\hat{\boldsymbol{r}}\cdot\boldsymbol{v}=-\frac{I(\boldsymbol{r},t+\Delta t)-I(\boldsymbol{r},t)}{\Delta t},
	\label{eq:DOFE}
\end{equation}

where $\Delta \boldsymbol{r}$ represents the inter-pixel displacement along the selected direction. Rearranging Eq.~\ref{eq:DOFE} provides an explicit directional velocity estimate:

\begin{equation}
	\boldsymbol{v}=-\frac{ \Delta\boldsymbol{r} }{\Delta t}\cdot\frac{I(\boldsymbol{r},t+\Delta t)-I(\boldsymbol{r},t)}{I(\boldsymbol{r}+\Delta \boldsymbol{r},t)-I(\boldsymbol{r},t)}\Delta\hat{\boldsymbol{r}}.
\end{equation}

To streamline computation, we apply binary quantization to the temporal difference term (retaining only its sign), preserving motion polarity while reducing sensitivity to intensity scaling. Leveraging the brightness constancy assumption in Eq.~\ref{eq:BCA}, we rewrite the denominator to couple intensity changes across neighboring pixels over time:

\begin{equation}
	\boldsymbol{v}=-\frac{ \Delta\boldsymbol{r} }{\Delta t}\cdot\frac{I(\boldsymbol{r},t+\Delta t)-I(\boldsymbol{r},t)}{I(\boldsymbol{r}+\Delta \boldsymbol{r},t)-I(\boldsymbol{r}+\Delta \boldsymbol{r},t+\Delta t)}\Delta\hat{\boldsymbol{r}}.
\end{equation}

For small $\Delta t$, intensity changes are minimal, and neighboring pixels typically maintain similar brightness. Under these conditions, ratio normalization and logarithmic approximation yield a more stable expression:

\begin{equation}
	\begin{aligned}
		\boldsymbol{v}&=-\frac{ \Delta\boldsymbol{r} }{\Delta t}\cdot\frac{\frac{I(\boldsymbol{r},t+\Delta t)}{I(\boldsymbol{r},t)}-1}{\frac{I(\boldsymbol{r}+\Delta \boldsymbol{r},t)}{I(\boldsymbol{r}+\Delta \boldsymbol{r},t+\Delta t)}-1}\cdot\frac{I(\boldsymbol{r},t)}{I(\boldsymbol{r}+\Delta \boldsymbol{r},t+\Delta t)}\Delta\hat{\boldsymbol{r}}\\
		&\approx-\frac{ \Delta\boldsymbol{r} }{\Delta t}\cdot\frac{\ln I(\boldsymbol{r},t+\Delta t)-\ln I(\boldsymbol{r},t)}{\ln I(\boldsymbol{r}+\Delta \boldsymbol{r},t)-\ln I(\boldsymbol{r}+\Delta \boldsymbol{r},t+\Delta t)}\Delta\hat{\boldsymbol{r}}.
	\end{aligned}
	\label{eq:eq7}
\end{equation}

Eq.~\ref{eq:eq7} expresses logarithmic intensity changes at neighboring locations over $\Delta t$. In event-based sensing, these correspond to ON/OFF responses indicating brightness change polarity at each pixel. Thus, motion estimation reduces to comparing ON/OFF responses between adjacent pixels along $\Delta\hat{\boldsymbol{r}}$, eliminating full-precision gradient computation.

This comparison-based approach mirrors biological visual processing. As shown in Figure~\ref{fig:1}(a), retinal bipolar cells convert photoreceptor outputs into distinct ON and OFF pathways, while direction-selective cells integrate these inputs across space and time for motion direction sensitivity. Similarly, our formulation utilizes relative ON/OFF responses from neighboring pixels to infer motion direction locally and distributedly. The reliance on relative consistency between neighboring changes inherently suppresses spurious noise events, as isolated single-pixel fluctuations typically fail to satisfy the required neighbor relationship.

\section{Bionic Circuit Design of OFS}

Based on the optical flow formulation in Eq.~\ref{eq:eq7}, the OFS employs an event-driven circuit implementation. As depicted in Fig.~\ref{fig:arch}(a), the system integrates two key elements: a pixel array for local optical flow computation, and an address-event interface circuit (\rev{OF-AER}) for asynchronous data handling. The pixel array features a dense 2D arrangement of OFS pixels, each processing brightness variations to compute local optical flow. Direct transmission of raw pixel data is infeasible due to quadratic interface scaling with array size, hindering on-chip scalability. Instead, optical flow signals are encoded as sparse events. The \rev{OF-AER} circuit then asynchronously collects these events and packages them into output packets containing spatial coordinates, temporal data, and motion vectors.

\subsection{OFS Pixel Circuit}

The schematic of a single OFS pixel appears in Figure~\ref{fig:arch}(b), where the pixel circuit locally extracts optical flow at the sensing site. As derived in \rev{Eq.~\ref{eq:eq7}}, optical flow is determined by comparing ON and OFF event responses from adjacent DVS pixels, per Eq.~\ref{eq:eq7}. DVS designs fall into two categories: standard CMOS-based architectures using photodiodes and novel optical memristor-based approaches, shown in Figure~\ref{fig:arch}(c). Optical memristor-based DVS improves area efficiency by fabricating photosensitive devices in upper metal layers, avoiding silicon substrate usage, enabling higher pixel density and better area utilization. However, optical memristor devices typically exhibit reduced photosensitivity and less ideal sensing performance relative to photodiodes, while requiring more complex resistance readout schemes.

\begin{figure}[b]
	\centering
	\includegraphics[width=\columnwidth]{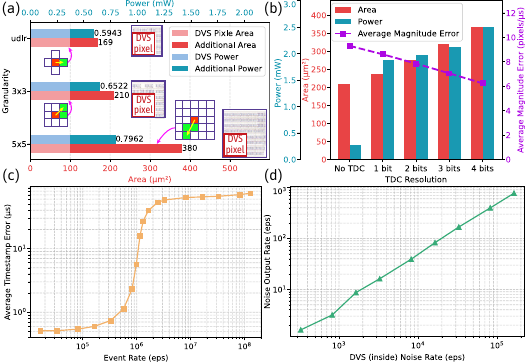}
	\caption{(a) Pixel-interaction granularity trade-off. (b) TDC resolution trade-off. (c) ATE of OFS vs. event rate. (d) OFS noise output rate vs. DVS (inside) noise rate.}
	\label{fig:granularity}
\end{figure}

An optical flow event at an OFS pixel is defined by temporal correlation between its OFF event and a neighboring pixel's ON event. \rev{In the design of pixel circuits, we only consider the ON-OFF event pairs between adjacent pixels.} When brightness decreases at a pixel, it generates an OFF event, followed by an ON event at a neighboring pixel after a brief delay as the moving edge arrives. The pixel's OFF event is broadened into a longer pulse, defining a temporal window $\Delta t$ (maximum transfer time). The broadened OFF signal combines with neighboring ON signals in an AND gate; if any neighboring pixel's ON event occurs within $\Delta t$, the AND output indicates an optical flow vector. A latch captures and holds this narrow pulse output. Concurrently, a TDC measures the time difference between the pixel's OFF event and a neighboring pixel's ON event, determining the optical flow magnitude. \rev{Since the optimal maximum transfer time for general scenarios is on the order of milliseconds, a simple combination of a latch and a counter (as shown in Figure~\ref{fig:arch}(b)) is employed as the TDC. The latch within the TDC also serves to provide broadened pulse. Consequently, the additional in-pixel computing circuits are entirely digital, making them significantly more resilient to transistor mismatch and process variations than analog solutions. Furthermore, full timing diagrams from our 1000-times Monte Carlo simulations are illustrated in Figure~\ref{fig:arch}(d). The digital design methodology facilitates the superposition of 1000-times Monte Carlo simulation runs, yielding a uniform waveform.} TDC resolution directly affects optical flow magnitude precision, though higher resolution increases area and power, as shown in Figure~\ref{fig:granularity}(b). Balancing flow accuracy, area, and energy, a 2-bit TDC resolution is chosen as optimal.

Each direction has an independent AND-latch path. Given that optical flow at a location typically has one dominant direction, each OFS pixel generates at most one optical flow vector before reset. OR operations combine detection results from all directions, generating a pixel-level event signal when flow is detected in any direction. Inter-pixel connectivity granularity determines motion information capture, with more connections providing richer spatial context but increasing area and power. Figure~\ref{fig:granularity}(a) shows granularity-area-power relationships for 4-connectivity, $3\times3$, and $5\times5$ neighborhoods. Simulations reveal area and power increase monotonically with granularity, leading to the $3\times3$ neighborhood as the optimal choice. \rev{It should be emphasized that, in principle, the $3 \times 3$ neighborhood does not result in missed detections. Given that the latency of DVS pixels (on the microsecond scale) is significantly shorter than the maximum transfer time $\Delta t$ (on the millisecond scale), the DVS pixels can asynchronously capture all ON/OFF events for objects traversing multiple pixels within $\Delta t$. Consequently, the OFS generates a continuous sequence of concatenated optical flow vectors.}

At array level, row event signals combine pixel-level events via OR gates, indicating optical flow activity and triggering OF-AER readout. To minimize pixel area, optical flow direction encoding is deferred to the bus stage. Post-data transfer, \texttt{row\_clr} resets row latches for the next cycle.

\subsection{Address-Event Interface Circuit: OF-AER}

The OF-AER circuit asynchronously reads optical flow vectors via row-wise processing, unlike conventional pixel-level AER. This reduces clock frequency demands, lowering power while maintaining throughput. Upon any pixel detecting flow, \texttt{row\_event} triggers an arbiter to grant exclusive bus access per row. An encoder generates the row address, decoded to \texttt{row\_load} for transferring direction and time-difference data to the bus. After one cycle, \texttt{row\_clr} clears the active row. The shift register serializes flow data during enable, using a clock frequency proportional to the array's column count. A synchronized column counter generates output addresses, while serialized data is buffered in a FIFO for downstream processing.

\section{Experiments}
\label{sec:eval}

\subsection{Configuration}

\begin{table*}[t]
	\centering
	\caption{Comparison table.}
	\begin{tabular*}{\textwidth}{@{\extracolsep{\fill}}p{0.22\textwidth}|p{0.22\textwidth}|p{0.22\textwidth}|p{0.22\textwidth}}
\toprule
		Sensor Design & DVS \cite{8778050} & OFS with CMOS DVS & OFS with optical memristor DVS \cite{https://doi.org/10.1002/adma.202511411}\\
\midrule
		Output Data Type & Dynamic Pixel Event & \multicolumn{2}{c}{Optical Flow Vector}\\
		\midrule
		Technology&65~nm&\multicolumn{2}{c}{65~nm}\\
		\midrule
		Sensor size &2~mm $\times$ 2~mm (1$\times$)&2.9~mm $\times$ 2.7~mm \textcolor{Bittersweet}{(1.96$\times$)} & 2.56~mm $\times$ 2.4~mm \textcolor{Bittersweet}{(1.54$\times$)}\\
		\midrule
		Resolution$^1$&132 $\times$ 104 (1$\times$)&\multicolumn{2}{c}{128 $\times$ 128 \textcolor{ForestGreen}{(1.19$\times$)}}\\
		\midrule
		Event Rate$^2$&388~keps$^3$ (1$\times$)&\multicolumn{2}{c}{100~keps$^4$ \textcolor{ForestGreen}{(0.26$\times$)}}\\
		\midrule
		Data Bandwidth Requirement &2.33~MB/s (1$\times$)&\multicolumn{2}{c}{700~KB/s$^5$ \textcolor{ForestGreen}{(0.30$\times$)}}\\
		\midrule
		Sensor Power$^6$ &4.9~mW (1$\times$)&7.4~mW \textcolor{Bittersweet}{(1.51$\times$)}&3.3~mW$^7$ \textcolor{ForestGreen}{(0.67$\times$)}\\
        \midrule
		Additional Processing Unit&FPGA \cite{9727106}&\multicolumn{2}{c}{Not Needed \textcolor{ForestGreen}{\checkmark}}\\
		\midrule
		Optical Flow Sense \& Compute Power&1~W (1$\times$)&7.4~mW \textcolor{ForestGreen}{(0.0074$\times$)}&3.3~mW$^7$ \textcolor{ForestGreen}{(0.0033$\times$)}\\ 
		\midrule
		Generation Latency/Event&1.1~$\upmu$s (1$\times$)&1~$\upmu$s \textcolor{ForestGreen}{(0.91$\times$)}&10~$\upmu$s \textcolor{Bittersweet}{(10$\times$)}\\
\bottomrule
		\multicolumn{4}{l}{\footnotesize{$^1$Resolution can be larger at the cost of sensor die area.}}\\
		\multicolumn{4}{l}{\footnotesize{$^2$Event rates are calculated based on the same scenario. Lower}}\\
		\multicolumn{4}{l}{\footnotesize{event rates are better for the purpose of effective motion data compression. $^3$Events are DVS pixel events. $^4$Events refer to}}\\
		\multicolumn{4}{l}{\footnotesize{optical flow vector events. $^5$Not all DVS events form optical flow vectors; we try our best to align the event numbers for a}}\\
		\multicolumn{4}{l}{\footnotesize{fair comparison. $^6$All powers are evaluated under the maximum event rate. $^7$The row scan time of the optical memristor-}}\\
		\multicolumn{4}{l}{\footnotesize{based OFS is set to 10~$\upmu$s to match the response time of optical memristors.}}\\
	\end{tabular*}
	\label{table:1}
\end{table*}

We evaluate the proposed OFS in this section. Both the OFS pixel array and the OF-AER readout circuitry are implemented using a commercial TSMC 65~nm CMOS process design kit. Power consumption is estimated through circuit-level SPICE simulations, while the area is obtained from the pre-silicon layout of the complete OFS design. The performance data of the optical memristor-based OFS are estimated based on the results reported in~\cite{https://doi.org/10.1002/adma.202511411}, where the read voltage $V_{\text{read}}$ is set to 0.1~V.

In addition to hardware-level metrics, the optical flow estimation behavior of the OFS is evaluated at the algorithmic level using the publicly available DVS09 dataset~\cite{4444573,Delbruck_2008} \rev{and MVSEC dataset~\cite{8288670}}. To ensure consistency between hardware and software evaluations, the same optical flow computation model implemented in the OFS circuitry is reproduced in Python for functional simulation and performance analysis.

\begin{figure}[!b]
	\centering
	\includegraphics[width=\columnwidth]{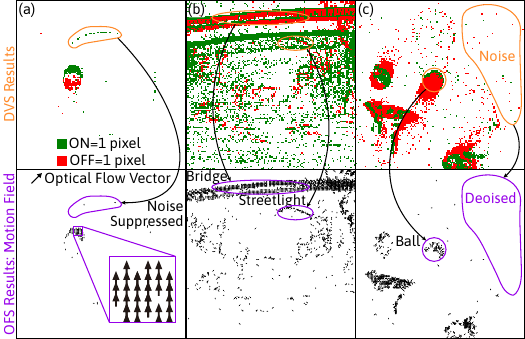}
	\caption{DVS sensor output data vs. OFS sensor output data: (a) Fast dot. (b) Freeway. (c) Juggling.}
	\label{fig:results}
\end{figure}

\subsection{Function Verification}

To validate the functional correctness and general applicability of the proposed OFS, simulation-based experiments are conducted on multiple representative dynamic vision scenarios. Example results are presented in Fig.~\ref{fig:results}. Three DVS sequences \rev{in DVS09 dataset} with distinct motion characteristics are selected for evaluation:

\noindent(a) \textit{Fastdot}, which contains rapidly rotating dot at 12000 rpm, used to verify the OFS response to high-speed local motion. The event rate of
this scenario is approximately 573~kilo events per second (keps).

\noindent(b) \textit{Freeway}, a driving scenario featuring vehicle motion, employed to assess the ability of the OFS to capture large-scale and fast translational motion, with an event rate of approximately 100~keps.

\noindent(c) \textit{\rev{Juggling}}, which includes complex and irregular motion patterns, used to evaluate robustness under non-uniform dynamics.

\rev{Concurrently, we evaluated the proposed algorithm on the more modern MVSEC dataset. Across various scenarios, the Average Endpoint Error (AEE) ranged from 159.73 pixels per second (px/s) to 274.28 px/s. In the indoor\_flying3 sequence, the AEE was 179.51 px/s, which underperforms compared to the results reported in \cite{9727106}. This discrepancy is attributed to the precision loss incurred by the quantization in Eq.~\ref{eq:eq7}, as well as the 2-bit quantization of the time difference.}

Across all scenarios, the OFS successfully detects motion direction and magnitude, demonstrating correct operation under diverse spatiotemporal conditions.

\subsection{Event Rate}

The ability of the OFS to sustain high optical flow activity is fundamentally constrained by the throughput of the OF-AER readout architecture. Unlike pixel-level readout schemes, the proposed design aggregates flow events at the row level, which introduces a bounded service rate determined by the row scanning speed. Specifically, the hardware permits the processing of one active row per microsecond, imposing a deterministic upper limit on the achievable event throughput.

When optical flow activity is sparse or moderately dense, row requests are rarely concurrent, and event readout closely follows event generation. Under such conditions, the temporal discrepancy between the actual occurrence of a flow event and its reported timestamp remains negligible. This behavior is reflected in Fig.~\ref{fig:granularity}(c), where the average timestamp error (ATE) remains negligible for event rates below 1~Meps.

As the event rate increases beyond this regime, multiple rows may assert readout requests simultaneously. Since these requests must be serialized by the arbiter, later rows experience increasing readout latency. This effect manifests as a gradual rise in ATE once the event rate exceeds 1~Meps. At the architectural limit, the OFS supports a peak output throughput of 128~Meps, corresponding to a configuration of 128 rows scanned at 1~$\upmu$s per row. Even at this extreme operating point, the measured ATE is approximately 74.3~$\upmu$s. This result demonstrates that the proposed OFS maintains bounded and predictable latency under ultra-high event rates, confirming the suitability of the OF-AER design for real-time, ultra-high-speed visual sensing applications.

\subsection{Noise Impact}

Event-based vision systems experience motion estimation degradation due to sensor noise-induced false events. Conventional DVS sensors typically exhibit intrinsic noise levels of 0.1--10~Hz/pixel~\cite{hu2021v2evideoframesrealistic}. We evaluate the noise robustness of our \rev{OFS} by analyzing how pixel-level DVS noise propagates through the optical flow extraction process. Figure~\ref{fig:granularity}(d) demonstrates the relationship between noisy input events and noise-induced flow vectors, with synthetic noise injected into the event stream and flow vectors generated exclusively under pure noise conditions identified as noise outputs. The OFS achieves noise suppression by a factor of 188$\times$ to 259$\times$ relative to input noise levels. This suppression emerges inherently from the optical flow computation architecture, which requires temporally correlated positive and negative events from spatially adjacent pixels within a limited time window. Isolated noise events, lacking spatial correlation, fail to meet this requirement and are naturally rejected, enhancing signal quality without additional computational or energy costs.

\subsection{Comparison to Related Works}

As summarized in Table~\ref{table:1}, unlike conventional DVS sensors that output pixel events and require external processing, the proposed OFS directly outputs optical flow vectors at the sensor level. This significantly reduces the event rate and data bandwidth by 3.8$\times$ and 3.3$\times$, respectively. By eliminating the need for FPGA-based post-processing, the OFS reduces the overall optical flow sensing and computing power by 135$\times$ to 303$\times$. Compared to the CMOS-based OFS, the optical memristor-based OFS achieves a 1.27$\times$ reduction in area and 2.31$\times$ in power consumption. The reduction in area is attributed to the use of optical memristors, which save valuable silicon area, while the decrease in power consumption results from the increased row scan time. However, while the CMOS-based OFS maintains microsecond-level latency, the optical memristor-based OFS exhibits a latency of approximately 10~$\upmu$s due to the \rev{light-induced response delay} of optical memristor devices. \rev{While both CMOS-based and optical memristor-based OFS are capable of performing in-pixel optical flow computation, the optical memristor-based OFS achieves significantly lower power consumption under identical scenarios and event rates.}

\section{Conclusion}

We propose a bio-inspired OFS that performs event-driven pixel-level optical flow computation. Experiments demonstrate over 303$\times$ power reduction with microsecond latency and approximately $3.3\times$ bandwidth saving compared to conventional DVS+FPGA. The OFS combines biological principles with efficient hardware for significant gains in power efficiency, latency, and bandwidth, suiting low-power ultra-high-speed vision applications.

\section{Acknowledgments}

This work has been supported by the National Key R\&D Program of China (2023YFB4502200), Guangdong Provincial Key Laboratory of In-Memory Computing Chips (2024B1212020002), Shenzhen Science and Technology Program (JCYJ20241202125907011), Beijing Natural Science Foundation (L234026, L257010), National Natural Science Foundation of China (92164302, T2350006, 92364102), and Financial Support for Outstanding scientific and technological innovation Talents Training Fund in Shenzhen. This work has been supported by the New Cornerstone Science Foundation. This work is sponsored by Beijing Nova Program.

\bibliographystyle{plainnat}
\bibliography{ref}

\end{document}